\documentclass[12pt,preprint]{aastex}

\newcommand{\hii}{\mbox{H{\sc~ii}}}

\begin{document}

\title{Infrared Observations of the Candidate LBV 1806-20 \& Nearby Cluster Stars\altaffilmark{1,2,3}}

\author{S.S. Eikenberry\altaffilmark{4,5,6}, K. Matthews\altaffilmark{7},
J.L. LaVine\altaffilmark{4}, M.A. Garske\altaffilmark{5,8}, D. Hu\altaffilmark{5},
M.A. Jackson\altaffilmark{5}, S.G. Patel\altaffilmark{4,5},
D.J. Barry\altaffilmark{5}, M.R. Colonno\altaffilmark{5},
J.R. Houck\altaffilmark{5}, J.C. Wilson\altaffilmark{5,9}, S. Corbel\altaffilmark{10}, J.D. Smith\altaffilmark{6,11}}

\altaffiltext{1}{Based on data obtained with the Cerro Tololo
Interamerican Observatory 4-meter telescope operated by NOAO.  NOAO is
operated by the Association of Universities for Research in Astronomy
(AURA), Inc. under cooperative agreement with the National Science
Foundation.}

\altaffiltext{2}{Based on data obtained at the Palomar Observatory
200-inch Telescope, which is operated by the California Institute of
Technology, the Jet Propulsion Laboratory, and Cornell University.}

\altaffiltext{3}{This publication makes use of data products from the
Two Micron All Sky Survey, which is a joint project of the University
of Massachusetts and the Infrared Processing and Analysis
Center/California Institute of Technology, funded by the National
Aeronautics and Space Administration and the National Science
Foundation.}

\altaffiltext{4}{Department of Astronomy, University of Florida, Gainesville, FL  32611}

\altaffiltext{5}{Astronomy Department, Cornell University, Ithaca, NY 14853}

\altaffiltext{6}{Visiting Astronomer, Cerro Tololo Interamerican Observatory}

\altaffiltext{7}{Physics Department, California Institute of Technology, Pasadena, CA  91125}

\altaffiltext{8}{Physics Department, Northwest Nazarene University, Nampa, ID 83686}

\altaffiltext{9}{Department of Astronomy, University of Virginia}

\altaffiltext{10}{Universit\'e Paris VII \& Service d'Astrophysique, CEA Saclay, F-91191 Gif sur Yvette, France}

\altaffiltext{11}{Steward Observatory, University of Arizona, 933 N. Cherry Ave., Tucson, AZ 85721}

\begin{abstract}

	We report near-infrared photometry, spectroscopy, and speckle
 imaging of the hot, luminous star we identify as candidate LBV
 1806-20 \footnote{We note that this star has not been observed to
 undergo a major LBV outburst yet, and thus some may not consider it
 to be an LBV (though it is undoubtedly luminous, blue, and variable,
 and shares many spectral characteristics with {\it bone fide} LBV
 stars, as we show here).  Hereafter, to avoid awkward phrasing, we
 refer to the candidate LBV as LBV 1806-20 as is done in the
 literature.} .  We also present photometry and spectroscopy of 3
 nearby stars, which are members of the same star cluster containing
 LBV 1806-20 and SGR 1806-20.  The spectroscopy and photometry show
 that LBV 1806-20 is similar in many respects to the luminous ``Pistol
 Star'', albeit with some important differences.  They also provide
 estimates of the effective temperature and reddening of LBV 1806-20,
 and confirm distance estimates, leading to a best estimate for the
 luminosity of this star of $> 5 \times 10^6 \ L_{\odot}$.  The nearby
 cluster stars have spectral types and inferred absolute magnitudes
 which confirm the distance (and thus luminosity) estimate for LBV
 1806-20.  If we drop kinematic measurements of the distance ($15.1
 ^{+1.8}_{-1.3}$ kpc), we have a lower limit on the distance of $>9.5$
 kpc, and on the luminosity of $>2 \times 10^6 \ L_{\odot}$, based on
 the cluster stars.  If we drop both the kinematic and cluster star
 indicators for distance, an ammonia absorption feature sets yet
 another lower limit to the distance of $>5.7$ kpc, with a
 corresponding luminosity estimate of $>7 \times 10^5 \ L_{\odot}$ for
 the candidate LBV 1806-20. Furthermore, based on very high
 angular-resolution speckle images, we determine that LBV 1806-20 is
 not a cluster of stars, but is rather a single star or binary system.
 Simple arguments based on the Eddington luminosity lead to an
 estimate of the total mass of LBV 1806-20 (single or binary)
 exceeding $190 \ M_{\odot}$.  We discuss the possible uncertainties
 in these results, and their implications for the star formation
 history of this cluster.

\end{abstract}

\keywords{stars: early type --- stars: emission-line --- stars: Wolf-Rayet --- stars: supergiants --- infrared: stars --- open clusters and associations: general}

\section{Introduction}

	Mounting evidence gathered in recent years indicates that
stars may be formed with masses much greater than previously thought
possible.  The hot luminous ``Pistol Star'' near our Galaxy's center,
for instance, has an estimated mass of $>150 \ M_{\odot}$
\citep{Figer98}, and the stars R136a1 and R136a2 in the Large
Magellanic Cloud each have masses of $140-155 M_{\odot}$
\citep{Massey}.  Since the luminosities of such massive stars exceed
$10^6 \ L_{\odot}$, relatively small populations of these stars can
dominate the power output of their host galaxy during their lifetimes.
Furthermore, their deaths may spread chemically-enriched material into
the galaxy and leave behind black holes as remnants.  The death events
may be responsible for gamma-ray bursts in the ``collapsar'' scenario
(e.g.  \citet{price}; \citet{hjorth}; \citet{MWH}), and the relatively
large remnant black holes may also explain the so-called
``intermediate-mass'' black holes currently being discovered in nearby
galaxies (e.g. \citet{IMBH} and references therein).  Thus, probing
the upper limit for stellar mass has an important impact on our
understanding of a wide range of astrophysical phenomena, including
the chemical evolution of matter in our Galaxy and external galaxies,
the history of galaxy and structure formation in the Universe, the
formation of black holes in their dying supernova events, and possibly
the origin of gamma-ray bursts via collapsars.

	We report here new observations of a luminous star we identify
as LBV 1806-20.  This star was first identified as a potential
counterpart to the soft gamma-ray repeater SGR 1806-20 \citep{Shri},
with high near-infrared brightness ($K = 8.4$ mag) despite significant
absorption from interstellar dust in the Galactic Plane.  Subsequent
moderate-resolution ($R \sim 700$) infrared spectroscopy revealed it
to be a candidate luminous blue variable (LBV) star and one of the
most luminous stars in the Galaxy, with $L > 10^6 \ L_{\odot}$
(\citet{vK95} ; \citet{Corbel}).  This star is known to lie at the
brightness peak of the radio nebula G10.0-0.3 (\citet{Shri93};
\citet{Vasisht}).  However, the revised Inter-Planetary Network (IPN)
localization of SGR 1806-20 indicated that the SGR was significantly
offset from the position of the candidate LBV star and the coincident
core of G10.0-0.3 \citep{Hurley}.  Recent {\it Chandra} and infrared
observations confirm that the SGR lies $\sim 12 \arcsec$ away from the
candidate LBV (\citet{kaplan}; \citet{eiken01}).  In addition,
\citet{gaensler} show that G10.0-0.3 is not a supernova remnant at
all, but is a radio nebula powered by the tremendous wind of the
candidate LBV star at its core.  The apparent conundrum presented by
this scenario -- why we find two such rare objects so close to each
other on the sky without any apparent physical connection -- has been
resolved by the observations of \citet{Fuchs} who showed that LBV
1806-20 is a member of a cluster of massive stars, and of
\citet{eiken01} who showed that SGR 1806-20 appears to be a member of
the same cluster.  Thus, while these two rare objects are not
identical, they are related through common cluster membership.
Interestingly, \citet{vrba} show that one of the other SGRs in the
Milky Way may also be associated with a massive star cluster.

The distance to the candidate LBV, SGR, and their associated cluster
has also been a subject of some discussion in the literature.  Initial
studies of CO emission from molecular clouds towards this line of
sight and the detection of $\rm NH_3$ absorption against the radio
continuum indicated an extinction of $A_V = 35 \pm 5$ mag and a best
estimate for the distance of $14.5 \pm 1.4$ kpc \citep{Corbel} --
again confirming its status as one of the most luminous stars known.
However, \citet{blum01} present infrared spectra of members of a
cluster in the nearby \hii \ region G10.2-0.3, which is also part of
the (apparent) W31 giant molecular cloud complex containing G10.0-0.3,
and find an apparently conflicting distance of $\sim 3.4 \pm 0.6$ kpc.
This apparent conflict has also been recently resolved by \citet{CE},
who present higher-resolution millimeter and infrared observations of
G10.0-0.3 and G10.2-0.3.  They find that W31 is actually resolved into
at least 2 components along the line-of-sight, with one component at
$d \sim 4$ kpc and with extinction $A_V \sim 15$ mag (in excellent
agreement with the \citet{blum01} observations of G10.2-0.3) and
another component at a (refined) distance of $d = 15.1^{+1.8}_{-1.3}$
kpc and with $A_V = 37 \pm 3$ mag.  The radial velocity of LBV 1806-20
matches both of these components, but the $NH_3$ absorption towards
the core of G10.0-0.3 (and thus LBV 1806-20) due to a cloud at $d =
5.7$ kpc unambiguously places the star in the ``far'' component of
W31.  In addition, both the infrared extinction towards LBV 1806-20
(\citet{vK95}; see also below) and the X-ray absorption column towards
SGR 1806-20 (\citet{eiken01}; \citet{mereghetti}) match the expected
extinction towards the ``far'' component of W31 and differ from that
of the ``near'' component of \citet{blum01} by $\sim 15$ mag, thus
confirming the association of the candidate LBV, SGR, and associated star
cluster with the ``far'' component of W31.

	In order to further investigate this intriguing object, we
obtained near-infrared images and spectra of LBV 1806-20 and several
nearby stars.  In Section 2, we describe these observations and their
reduction.  In Section 3, we describe the analysis of the resulting
data, including the spectral types of the candidate LBV and cluster stars,
refined analyses of the reddening, confirmation of the distance
estimate of \citet{CE}, and the resulting luminosity estimate for LBV
1806-20.  In Section 4, we discuss the uncertainties in these
measurements and their implications for our understanding of the
formation and evolution of extremely massive stars and the birth
environment of SGRs.  Finally, in Section 5 we present our
conclusions.

\section{Observations and Data Reduction}

\subsection{CTIO -- July 2001}

	We used the Ohio State InfraRed Imaging Spectrograph (OSIRIS)
intrument \citep{Depoy} and f/14 tip-tilt secondary on the
Cerro-Tololo Inter-American Observatory (CTIO) 4-meter telescope on
July 5-6, 2001 to observe LBV 1806-20 and nearby stars.  In the OSIRIS
imaging mode, we obtained $J-$, $H-$, and $K-$band images with a
$0.161 \arcsec \ {\rm pixel^{-1}}$ plate scale.  While conditions were
non-photometric due to high clouds, the seeing conditions were
acceptable when the transparency allowed observations -- using the
tip-tilt secondary, typical images had full-widths at half-maximum
(FWHM) of $\sim 0.6-0.7 \arcsec$.  In each band, we obtained a set of
9 images in a $3 \times 3$ raster pattern with a $10 \arcsec$ offset
between images.  We then subtracted dark frames from each image,
divided the result by its own median, and then median-combined the
resulting images into a normalized sky frame.  We then subtracted the
dark frame and a scaled version of the sky frame from each of the 9
images, and divided the result by a dome flat image.  We shifted each
of the 9 frames to a common reference position and added them to give
a final summed image in each band.  Figure 1 shows a composite 3-color
image of the field of LBV 1806-20 in the following near-infrared
bands: $J = 1.25 \mu {\rm m}$ (blue); $H = 1.65 \mu {\rm m}$ (green);
$K = 2.2 \mu {\rm m}$ (red).

	We also used the OSIRIS high-resolution spectroscopic mode to
obtain moderate resolution ($R = 1500$ for a 4-pixel slit) spectra of
LBV 1806-20 and three nearby stars (B, C, D in Figure 1).  This group
of stars, approximately $12 \arcsec$ west of LBV 1806-20 appears to be
a cluster of young, hot, luminous stars in the same molecular cloud
as, and including, the candidate LBV star \citep{Fuchs}.  At the
estimated distance of 15.1 kpc \citep{CE}, this angular separation
corresponds to a physical distance of only $\sim 1$ pc between the
candidate LBV star and the center of the cluster, implying a common
origin.  Thus, we chose the three brightest stars near the center of
the association for spectroscopic observation.  We oriented the OSIRIS
slit to obtain spectra simultaneously of LBV 1806-20 and Star B, and
separately to obtain simultaneous spectra of Stars C and D.  We used
the high-resolution slit of OSIRIS (4 pixels $ = 0.67 \arcsec$), and
took $6 \times \ 120-$s exposures at positions offset along the slit.
After each set of 6 exposures, we obtained a set of 6 spectra of the
nearby G dwarf star HR6998, with approximately the same positions
along the slit.  We repeated this approach for 2 different grating
settings for each of the K and H bands, and one grating setting for
LBV 1806-20 in the J-band.  For both the science targets and the
G-star, we took each spectral frame, subtracted a dark frame from it,
divided by its own median, and median-combined the 6 resulting images
for a given grating tilt to create a normalized sky frame.  We then
subtracted a dark frame and a scaled sky frame from each image and
divided by a spectral dome flat.  We extracted spectra separately from
each processed image with a Gaussian weighting in the spatial
direction and tracking the curvature of the spectrum in the dispersion
direction.  We then divided each target spectrum by the G-star
spectrum with the nearest position on the slit to remove atmospheric
absorption bands, after interpolating over the Brackett absorption
features in the G-star spectrum.  We obtained a wavelength solution
separately for each spectrum using the OH lines nearby on the sky
(typical residuals $<20 \ {\rm km \ s^{-1}}$ in velocity space), and
used this to correct each spectrum to a common wavelength scale before
averaging the spectra to a single final spectrum.  For H and K bands,
we averaged the 2 spectra from the different grating tilts where they
overlapped.  We multiplied the result by a 5600K blackbody spectrum
(corresponding to the temperature of HR 6998), and de-reddened them
for $A_V = 29$ mag (see below for further discussion on the
reddening).  We present these spectra in Figures 2-6.

\subsection{HBO -- July 2001}

	Due to the non-photometric conditions during the CTIO
observations, we were unable to photometrically zeropoint-calibrate
the images obtained at that time.  Thus, we post-calibrated these
images using observations of stars in the field of LBV 1806-20 and
infrared standard stars on July 27, 2001 with the Hartung-Boothroyd
Observatory 0.65-meter telescope and its infrared array camera
\citep{Houck}, as well as with photometry from the Two-Micron All-Sky
Survey (2MASS).  We took 7 images of the field of LBV 1806-20 in each
band, with offsets of $\sim 15 \arcsec$ between images.  We then
processed the images in each band as described for the CTIO data
above.  We repeated this procedure on sets of 7 images of the UKIRT
standard FS26 in each band.  We extracted the flux in $\rm ADU / s$
from each processed image of FS26 individually, using the average as
the best estimate of the flux, and the standard deviation divided by
the square root of the number of exposures as the $1 \sigma$
uncertainty.  We then used the known magnitudes of this star to
calibrate similarly-derived flux measurements and uncertainties for
several bright stars in the field of LBV 1806-20.  Differential
photometry between these stars and the target stars provided the final
photometric measurements.  We used a photometric airmass solution
derived from measurements of a star observed for another program over
a range of airmasses throughout the same night.  We then verified this
photometry with the 2MASS photometry for J and H band (omitting K, as
2MASS uses the $K_s$ filter).  We present the resulting photometry for
the bright stars of Figure 1 in Table 1.

\subsection{Palomar -- July, 1999}

	We obtained high angular-resolution speckle imaging of LBV
1806-20 in June 1999 using the Palomar 5-meter telescope.  We used the
facility near-infrared camera D78 with reimaging fore-optics providing
a pixel scale of $0.036 \arcsec (\pm 0.001 \arcsec)$ per pixel to
obtain K-band images of the star with 0.125s exposures, effectively
freezing the effects of image motion due to atmospheric turbulence.
We observed LBV 1806-20 in 6 sets of 50s on-target integrations,
interleaved with observations of 2 nearby stars with similar
near-infrared brightnesses taken in an identical manner.  We used a
``shift-and-add'' technique to combine images of the individual stars
-- for each star (LBV 1806-20 and the 2 comparison stars) we shifted
the images to align the brightest speckle in each frame with other
frames and added the results.  We used these nearby stars to create
the model point spread function (PSF), which were seen to have FWHM of
$0.130 \arcsec$ -- near the telescope's diffraction limit of $0.110
\arcsec$ at this wavelength.  We then scaled the amplitude of this
model PSF and subtracted it from the shifted-and-added image of LBV
1806-20.  We also used Fourier filtering to remove a periodic diagonal
pattern due to clocking noise in the camera electronics.  Figure 7
shows an image of LBV 1806-20 and the resulting PSF-subtracted image.
While the residuals in the subtracted image are statistically
significant, they may be due to systematic effects, such as sub-pixel
errors in image registration or secular small-scale variations in the
speckle PSF, possibly due to low-level instability in the
time-averaged properties of the speckle halo.  For comparison, we also
show simulated PSF-subtracted images of a theoretical extended source
with a very small size ($0.06 \arcsec$ FWHM).  To simulate this
source, we took the comparison star PSF and convolved it with a 2-D
Gaussian profile with a width corresponding to the extended source
intrinsic width.  We then added noise corresponding to the photon
noise in the LBV 1806-20 image to the resulting simulated image, and
then subtracted a scaled version of the single-star PSF as above.
Note that the resulting simulations have residuals much greater than
those for the actual PSF-subtracted image.

\section{Analysis}

\subsection{Spectral Analysis}

\subsubsection{LBV 1806-20}

	Based on the spectra in Figures 2-4, we present measured
equivalent widths for emission and absorption lines from LBV 1806-20
in Table 2.  The line features evident in the K-band spectrum of LBV
1806-20 (HI, HeI, FeII, MgII, NaI) are very similar to those observed
in LBV stars such as AG Car \citep{Morris}, the Pistol Star in the
Quintuplet \citep{Figer98}, and the LBV candidate Star 362 also in the
Quintuplet \citep{Geballe} (and at lower resolution from this same
star in 1994 \citep{vK95}).  The equivalent widths of the FeII, MgII,
and NaI features are very close also, though the HI and HeI lines are
generally stronger in LBV 1806-20.  P Cygni also displays a similar
K-band spectrum \citep{Smith}, although with an apparent absence of
NaI emission.  In the H-band, LBV 1806-20 seems like a cross between P
Cygni and the Pistol Star, with 10 strong Brackett series lines and 9
FeII emission lines.  While the Pistol Star and AG Car exhibit similar
iron features, their Brackett series lines are much weaker than seen
in LBV 1806-20.  P Cygni, on the other hand, shows strong Brackett
series with weaker FeII.  Thus, while the spectrum of LBV 1806-20 does
not uniquely match any particular LBV, its properties are well within
the range exhibited by these stars.  Thus, we conclude that based on
its spectrum LBV 1806-20 is in fact a luminous blue variable
candidate, in confirmation of \citet{vK95}.

	The HeI $2.112 \mu$m absorption line provides important
information on the temperature of LBV 1806-20.  In 1994, the
equivalent width of this line was $1.8 \pm 0.4 \ {\rm \AA}$
\citep{vK95}, consistent with spectral classes O9-B2 of supergiant
stars \citep{Hanson}, indicating a surface temperature of 18000 K to
32000 K \footnote{We note that it is conceivable that the
temperature-EW relation for supergiants (which has rather large
scatter) does not apply to candidate LBV stars.  However, independent
temperature measurements for the Pistol Star \citep{Figer98} are
consistent with the temperatures derived from its HeI $2.112 \mu$m
absorption feature, indicating that this approximate relationship
seems to continue to hold for candidate LBV stars.}.  The large range
in temperature is due to the large scatter in this relationship
\citep{Hanson}.  By the time of the July 2001 observations, the
equivalent width of the line had dropped to $0.76 \pm 0.17 \ {\rm
\AA}$.  Changes in the spectral type to hotter (O6.5-O8.5) surface
temperature would produce such a reduction in the HeI absorption line
equivalent width \citep{Hanson}.  Moreover, at the same time, we see a
factor of $\sim 4-6$ increase in the equivalent width of emission
lines such as ${\rm Br} \gamma$ and HeI $2.058 \mu$m, which could
reflect an increase in the number of ionizing photons produced by the
star as it moved to higher temperatures.  However, such a temperature
change might also imply a significant brightening in the K-band (given
a constant photospheric radius), contrary to the observations.  In
fact, our K-band photometry for LBV 1806-20 is marginally fainter than
that of \citet{Shri} ($K = 8.4$ mag, with no quoted uncertainty).  We
can find an alternate solution which keeps the K-band brightness
constant while decreasing the HeI absorption and increasing the number
of ionizing photons, by assuming that the photospheric radius
decreases significantly as temperature increases.  Simple calculations
for a blackbody spectrum show that a change of temperature from
20,000K to 26,000K (roughly B1.5 to B0) accompanied by a $\sim 25\%$
decrease in photospheric radius could produce the observed behavior.
Such anti-correlated variations in temperature and radius are
characteristic of some LBV stars (\citet{HD96}; \citet{Morris}).
Alternately, the stronger lines might indicate an increase in the
stellar wind density absorbing more ultraviolet continuum photons and
converting them into increased line emission.  However, without
detailed models of the atmosphere of LBV 1806-20, we can only conclude
that it has a spectral type between O9 and B2.

\subsubsection{Star B}

	From the color of Star B in Figure 1, we can immediately see
that it differs significantly from the other stars in the field due to
its extreme redness ($J-K = 7.3$ mag versus $J-K < 6.0$ mag for other
stars).  In addition, the emission line spectrum (Figure 5) differs
significantly from LBV 1806-20, showing relatively weak $\rm Br
\gamma$ and strong blended emission from HeI, HeII, CIII, CIV, NIII,
and NIV.  This K-band spectrum is typical for late-type Wolf-Rayet
stars of the WC subclass, and comparison of the equivalent widths of
these lines with other similar stars \citep{FigerWR} gives a
classification of WC9.  The long wavelength excess in the continuum
indicates the presence of warm dust with a thermal continuum extending
into the near-infrared -- another common feature in WCL (late-type WC)
stars, and also explaining the very red colors of this object.  We
should note that only 26 such WCLd (dust emission) stars are known at
this time \citep{vdH}, making this a rather rare star.

\subsubsection{Stars C and D}

	The next brightest stars in the cluster center, C and D, have
similar colors to LBV 1806-20, with $J-K = 5.6$ mag and $J-K = 4.9$
mag, respectively.  In their K-band spectra, both stars show $\rm Br
\gamma$ and HeI $2.112 \mu$m absorption features.  They also seem to
show some sort of feature at the HeI $2.058 \mu$m line.  This could
either be due to a self-absorbed (nascent P-Cygni) line, or else due
to poor subtraction of a nearby strong OH sky line.  Based on their
strong HeI $2.112 \mu$m absorption lines, both of these stars would
seem to have spectral classes in the range from B0 to B3 supergiants
\citep{Hanson}.  However, in both of these absorption lines, as well
as the stronger of the Brackett series lines in the H-band spectra for
these stars, we see some evidence for absorption wings (Figure 6).
Fits to these lines give velocity widths of $>200 {\rm km \ s^{-1}}$
FWHM (after subtraction of the instrumental line widths), although
some of these lines are potentially blended (particularly Br$\gamma$
with nearby He lines).  Such broad absorption wings are typical of the
luminous blue hypergiants (luminosity class Ia+), and may indicate the
presence of a mass-losing wind \citep{deJager}.  Thus, we conclude
(for the time being) that Stars C and D are B0-B3 supergiant or
hypergiant stars.

\subsection{The Reddening towards LBV 1806-20}

	The colors of LBV 1806-20 and its spectral continuum shape
allow us to estimate the extinction towards LBV 1806-20.  For such a
hot star (as indicated by the HeI $2.112 \mu$m absorption feature),
the intrinsic $J-K$ color is nearly neutral, and the observed red
color of $J-K = 5.0 \ \pm 0.15$ mag corresponds to an extinction of
$A_V = 28 \pm 3$ mag (assuming the Rieke-Lebofsky reddening law
\citep{RiekeLebofsky}), matching the estimates based on CO
observations \citep{CE}.  (While a hypothetical infrared excess from
LBV 1806-20 would alter these conclusions, as we discuss below there
is good reason to believe this is not present).  Furthermore, the fact
that stars C and D in the nearby cluster have very similar $J-K$
colors (as can also be seen for many other cluster stars' colors in
Figure 1) indicates that the stellar cluster is indeed at the same
reddening and thus distance along the line of sight as LBV 1806-20,
confirming the physical association between them (see also
\citet{eiken01}).  In addition, the H and K bands are in the
Rayleigh-Jeans portion of the blackbody emission curve for such a hot
star (the reason for the neutral colors noted above).  Thus, we can
estimate the extinction towards LBV 1806-20 by de-reddening the
spectra until the continuum shape matches a Rayleigh-Jeans
distribution.  In this way, we obtain estimates of $A_V = 31 \pm 3$
mag from the H-band continuum and $A_V = 28 \pm 3$ mag from the K-band
continuum, with uncertainties dominated by $\sim 10\%$ uncertainty in
the spectrograph response shape over a given order.

For the $J-K$ measurement above, adoption of the Cardelli reddening
law \citep{Cardelli} gives a very small change in $A_V$ ($<0.4$ mag),
and the K-band spectral continuum shape is similar unaffected (change
of $<0.1$ mag).  In the H-band, the Cardelli law differs from
Rieke-Lebofsky more significantly, causing a difference of $\sim 1.9$
mag.  Taking this into account, we increase our uncertainty in that
measurement to $A_V = 28 \pm 4$ mag.  Combining all three, we adopt a
final estimate for the extinction of $A_V = 29 \pm 2$ mag towards LBV
1806-20.

\subsection{The Distance to LBV 1806-20}

The distance to LBV 1806-20 is a major subject of \citet{Corbel} and
\citet{CE}, and we refer the reader to those papers for detailed
discussion.  In summary, \citet{Corbel} and \citet{CE} use CO
spectroscopy to identify molecular clouds along the line-of-sight
towards LBV 1806-20, and use the cloud velocities to determine
kinematic distances to them.  The spectra of LBV 1806-20 presented
here provide important insights into the distance of the star.  From
the emission lines, we can measure a radial velocity of LBV 1806-20.
We selected the $\rm Br \gamma$ line as a velocity fiducial, as it is
the strongest line detection in the spectrum, and appears to be
relatively free from contamination due to blending with other strong
lines.  We fit a Gaussian profile to this line, finding no significant
residuals, and a centroid shifted from the atmospheric rest frame by
$-3 \pm 20 \ {\rm km \ s^{-1}}$, where residuals in the spectral
wavelength solution from atmospheric OH emission lines dominate the
largely systematic uncertainty.  After correcting for the Earth's
barycentric motion and the Solar System barycenter motion relative to
the local standard of rest, we determine a radial velocity for LBV
1806-20 of $v_{LSR} = 10 \pm 20 \ {\rm km \ s^{-1}}$.  Cross-checks of
this velocity determination with several other strong unblended lines
give consistent results for the velocity of LBV 1806-20.  This
velocity is important, as massive stars such as LBVs are a
kinematically ``cold'' population, and do not generally deviate
significantly in velocity from their parent molecular clouds.
Combined with the CO velocity maps of \citet{Corbel} and \citet{CE},
this velocity then confirms the association of LBV 1806-20 with
molecular clouds in W31 at kinematic distances of either $\sim 4$ kpc
or $\sim 15$ kpc, based on the Galactic rotation curve \citep{Fich}
and with the ambiguity being due to the near/far degeneracy of
kinematic distances.

\citet{CE} also present an $NH_3$ absorption spectrum
which uses the radio emission from LBV 1806-20 as the background
source, revealing strong absorption from a molecular cloud whose
velocity (in both $NH_3$ absorption and $CO$ emission) places it at a
near distance of $5.7$ kpc, setting this as the lower limit to the
distance of LBV 1806-20.  This observation eliminates the ``near''
distance as a possibility for LBV 1806-20, leaving only the ``far''
distance of $d = 15.1 ^{+1.8}_{-1.3}$ kpc.  As a ``sanity check'',
\citet{CE} go on to show that the observed reddening towards LBV
1806-20 combined with $CO$ observations of clouds along the line of
sight is consistent with the ``far'' distance and is inconsistent with
the ``near'' distance.

Based on these results, we adopt the distance determination of
\citet{CE} of $d = 15.1 ^{+1.8}_{-1.3}$ kpc.  We note that an
independent distance estimate for Star B (below) matches this estimate
and is also strongly inconsistent with the ``near'' distance noted
above.  We discuss this issue further in section 4.1.3.

\subsection{The Luminosity of LBV 1806-20}

	Combining the above measurements, we then arrive at luminosity
estimates for LBV 1806-20.  Taking its brightness of $K = 8.89 \pm
0.06$ mag and applying an extinction correction of $A_K = 3.2 \pm 0.2$
mag (corresponding to $A_V = 29 \pm 2$ mag and $A_K = 0.112 A_V$
\citep{RiekeLebofsky}) and a distance modulus of $15.9 \pm 0.2$ mag
(from $d = 15.1 ^{+1.8}_{-1.3}$ kpc), we arrive at an absolute K
magnitude of $M_K = -10.2 \pm 0.3$ mag.  The absolute visual magnitude
and bolometric luminosity of LBV 1806-20 from this number are
functions of the star's spectral class, and also may be affected by
free-free contributions to the K-band emission of LBV 1806-20.  As
discussed below, we do not believe that this contribution is large for
LBV 1806-20 (it is $< 0.1$ mag for the Pistol Star as well), and we
include it as an additional $0.1$ mag uncertainty in the lower bound
for the absolute K-band magnitude, which we now adopt to be $M_K =
-10.2 ^{+0.4} _{-0.3}$ mag.  For a spectral class of O9, the upper end
of the range, we have $V-K = -0.8$ mag to give an absolute visual
magnitude of $M_V = -11.0 ^{+0.4} _{-0.3}$ mag.  The bolometric
correction is $BC = -3.2 \pm 0.2$ mag, giving a bolometric magnitude
of $M_{bol} = -14.2 ^{+0.5} _{-0.4}$ mag, or a luminosity of $\sim 4
\times 10^7 \ L_{\odot}$.  At the low end of the temperature range (B2
spectral type), the corresponding values are $M_V = -10.6 ^{+0.4}
_{-0.3}$ mag, $M_{bol} = -12.0 ^{+0.5} _{-0.4}$ mag, or a luminosity
of $\sim 5 \times 10^6 \ L_{\odot}$.  We have plotted these
temperature/luminosity values in Figure 8, along with the
corresponding locations of other known extremely luminous stars.  Note
that even for the lower end of the possible temperature range for LBV
1806-20, it has a luminosity equal to or greater than that of the
famous LBV Eta Carina \citep{STIS} and the Pistol Star
\citep{Figer98}.  Thus, it seems that LBV 1806-20 may (marginally) be
the most luminous star currently known.

\subsection{The Absolute Magnitude and Distance of Star B}

Given the photometry, distance estimate, and reddening above, we can
also arrive at an abolute magnitude for Star B.  If we assume that the
reddening toward star B is identical to that for LBV 1806-20 and a
distance of 15.1 kpc, the absolute K-band magnitude for this star is
$M_K = -8.6 \pm 0.3$ mag.  Note that this is in excellent agreement
with the range of absolute magnitudes for WC9 stars in the Galactic
Center ($M_K = -8$ to $-11$ mag -- \citet{blum03}).  Furthermore, the
observed $H-K$ colors of star B are consistent with the assumed
reddening if the intrinsic colors are $H-K \sim 1.1$ mag, also in
excellent agreement with the observed range of instrinsic colors of
other WC9 stars ($H-K = 0.9$ to $1.6$ mag \citep{blum96}) \footnote{We
note that there is not always a one-to-one mapping between infrared
and optical spectral classifications of W-R stars, which might suggest
some uncertainty here.  However, we are using an infrared spectral
classification of Star B to compare its infrared properties to other
infrared-classified WC9 stars here.  Therefore, we conclude that this
comparison should be relatively free from any confusion due to
differing optical/infrared classifications.}.  Thus, the spectral
classification and photometry of star B confirm the distance and
reddening estimates used for the luminosity of LBV 1806-20 above.

Taking a slightly different approach, we can use the observed range of
intrinsic colors and absolute magnitudes for WC9 stars to constrain
the distance to star B and provide a completely independent
cross-check on the distance to the cluster of stars including LBV
1806-20.  The intrinsic color range of \citet{blum96} combined with
the observed $H-K = 2.96 \pm 0.08$ mag and the Rieke-Lebofsky
reddening law gives a range of $3.6 > A_K > 2.8$ mag.  Combining this
with the observed range of $M_K$ \citep{blum03} for Galactic Center
WC9 stars and the observed $m_K = 10.50 \pm 0.06$ mag gives a range
for the distance modulus to star B of $m_d = 14.9$ mag
(low-luminosity, intrinsically-blue, high-reddening) to $18.7$ mag
(high-luminosity, intrinsically-red, low-reddening).  Thus, a lower
limit on the distance of star B is $ d> 9.5$ kpc, assuming it is no
fainter than the intrinsically-faintest known WC9 star in the Galactic
Center (an assumption confirmed by the recent discovery of another,
even fainter WC9 star in this same cluster \citep{lavine}).  Note that
at an alternate assumed distance of $\sim 4$ kpc (consistent with the
G10.2-0.3 cluster of \citet{blum01}), star B would be 6 times less
luminous than any other known WC9 star, which seems very unlikely
(especially given the even fainter WC9 of \citet{lavine}).  Thus, this
provides yet another independent confirmation that these stars lie in
the ``far'' component of W31 at $d = 15.1 ^{+1.8}_{-1.3}$ kpc, as
opposed to the ``near'' component observed by \citet{blum01}.

\subsection{The Luminosities of Stars C \& D}

	The same analyses applied to LBV 1806-20 can also provide
luminosity estimates for stars C and D.  Based on their strong HeI
$2.112 \mu$m absorption lines, both of these stars have spectral
classes in the range from B0 to B3 \citep{Hanson}.  Following the
analyses for LBV 1806-20 (above), at the distance of 15.1 kpc, for a
B1 spectral class, they would have bolometric luminosities of $\sim
1.6 \times 10^6 \ L_{\odot}$ and $1.4 \times 10^6 \ L_{\odot}$,
respectively.  These luminosities are too high for ``normal''
supergiant stars of this spectral type.  However, these luminosities
are similar to those observed in early-B hypergiants (luminosity class
Ia+) \citep{deJager} (see Figure 8).  Furthermore, hypergiants are
distinguished from ``normal'' supergiants by the presence of broad
absorption lines indicative of mass-loss through a wind -- much as we
see in the $>200 {\rm km \ s^{-1}}$ linewidths of the Brackett and HeI
absorption features in Stars C and D (though, as noted above, line
blending may have a non-neglible effect for these lines at this
resolution).  Therefore, the luminosity and spectral features of Star
C and D appear to support their identification as class Ia+ blue
supergiants at the same distance and reddening as LBV 1806-20.

Furthermore, if we take the same reddening and distance used for LBV
1806-20, we derive absolute magnitudes of $M_K = -8.3 \pm 0.3$ mag and
$M_K = -8.0 \pm 0.3$ mag, respectively for stars C and D, without
reference to (potentially uncertain) bolometric correction.  We note
that there are several luminous stars in the Galactic Center with
similar effective temperatures (i.e. stars IRS 16SW and IRS 16NE --
\citet{najarro}) and absolute magnitudes of $M_K = -7.5$ to $-8.0$
mag, in excellent agreement with stars C and D.  While the IRS 16
stars are emission-line objects, and stars C and D are absorption line
objects, we can at least see that the absolute magnitudes of the
brightest B-type stars in this cluster are consistent with the
brightest B-type stars in the Galactic Center.  This supports at least
the {\it consistency} of the distance and reddening estimates used for
LBV 1806-20.  Finally, we note that in a cluster such as this, it
seems quite likely that some of the brightest stars are in fact
binaries, given the large fraction of binarity in high-mass stars.
Thus, by selecting the brightest stars in this cluster for
spectroscopy, we may be biasing ourselves towards binary stars with
apparent luminosity excess (see below for a more detailed discussion
of multiplicity in LBV 1806-20).

\section{Discussion}

\subsection{Caveats to the Luminosity Estimate for LBV 1806-20}

\subsubsection{Near-infrared excess}

	One issue for the above luminosity estimate is that LBV
1806-20 might exhibit a near-infrared excess of continuum emission due
to warm circumstellar dust (as Star B), free-free emission in the LBV
wind, or other processes.  Such an IR excess would interfere with our
luminosity estimate in several ways.  First, it would artificially
enhance the K-band brightness of the star, and thus directly cause us
to over-estimate the luminosity of LBV 1806-20.  Second, it would
artifically decrease the apparent equivalent width of the HeI $2.112
\mu$m absorption feature, causing us to mis-estimate the star's
temperature and thus its bolometric correction.  Finally, it would
redden the colors and spectral continuum shape of candidate LBV 1806-20, causing
us to over-estimate the reddening correction for the star.  Thus, we
see that the hypothetical presence of such an excess could
significantly alter our luminosity estimate.

	However, careful inspection of the observational results gives
no evidence for such an excess, and provides several indications
against its existence.  First of all, as we noted above and as can be
seen in Figure 1, there are several stars near LBV 1806-20 with very
similar $JHK$ colors.  If LBV 1806-20 has a significant near-IR
excess, then we must also postulate that the majority of the bright
stars in/near the cluster also have significant (and virtually
identical!) excesses.  This is extremely unlikely, if not positively
unphysical.  Second, the spectral continuum shape in {\it both} the H
and K bands is consistent with a reddened Rayleigh-Jeans distribution
for LBV 1806-20 (as well as Stars C and D).  On the other hand, the
spectral continuum shape in Star B clearly reveals its near-IR excess
(likely due to warm dust).  Therefore, for an excess to be present in
LBV 1806-20, it must extend smoothly over at least the entire H and K
bands, thereby significantly altering the observed colors of the star.
This conflicts with the observed color match with other cluster stars
noted above.

We also note that the LBV stars which most closely resemble LBV
1806-20 in their emission line spectra -- AG Car and the Pistol Star
-- do not show large free-free emission contributions in this
wavelength range.  The free-free contribution of the Pistol Star in
the K-band is negligible for models of the spectral energy
distribution \citep{Figer98} and estimated to be $< 0.1$ mag
(D. Figer, private communication), while in AG Car the inferred K-band
contribution is $\sim 0.2$ mag \citep{McGregor}.  While AG Car {\it
does} exhibit mid-IR excess emission, it does not contribute
significantly at wavelengths $\la 10 \mu$m.

Finally, we note that the supposed dilution of the HeI absorption
feature by a large (e.g. $>50$ \%) near-IR excess would imply an
intrinsic equivalent width $>4 \rm \AA$ (for an IR excess equal to the
blackbody continuum in K-band).  However, the temperature range we
infer for LBV 1806-20 is near the maximum strength of this line.  In
fact, {\it none} of the stars in the census of \citet{Hanson} have
equivalent widths $>3 \rm \AA$.  Thus, the presence of significant
dilution of this line by an IR excess would make its spectrum
inconsistent with any known type of star.

For all of these reasons, we conclude that LBV 1806-20 does not
exhibit any {\it significant} near-IR excess emission over that
expected for a reddened blackbody.  By ``not significant'', we mean
here that the contribution of any IR excess is not large compared to
the other uncertainties in our measurement of the star's luminosity
($\sim 0.4$ mag).  However, as noted in Section 3.4, we have added an
additional $0.1$ mag of uncertainty to the lower bound for LBV 1806-20,
based on a possible expectation of this level of free-free emission as
observed in the similar Pistol Star.

\subsubsection{Temperature}

	The issue of the precise temperature for LBV 1806-20 is a
difficult one.  At a crude level, the spectral continuum shape in H
and K bands gives us a firm lower limit of $\sim 12000K$, below which
we would see spectral curvature away from a Rayleigh-Jeans law,
contrary to the observations.  However, our primary indicator for the
temperature of the star is its HeI $2.112 \mu$m absorption feature.
The simple presence of this line also indicates an effective
temperature greater than $12000 K$ \citep{Hanson}.  Assuming the
relation found between temperature and equivalent width for supergiant
stars gives the temperature range from $18000K$ to $32000K$, as noted
above.  The case of the Pistol Star seems to confirm that this
relationship extends to some candidate LBV stars also \citep{Figer98}.

	The variability of this line in LBV 1806-20 unfortunately
complicates the matter.  As noted above, a change in the stellar
temperature to either greater or smaller values, could produce the
observed reduction in equivalent width.  It has been argued
\citep{HD96} that LBVs have essentially constant bolometric
luminosities, and that their apparent brightness variations are due to
anti-correlated radius/temperature variations induced by their
near-Eddington radiative instability.  Thus, a star can experience an
increase in temperature with a simultaneous decrease in radius,
keeping the bolometric luminosity constant but significantly altering
its apparent brightness at wavelengths near the peak of the blackbody
spectrum.  The observed decrease in the HeI line from LBV 1806-20
could represent such a temperature/radius change.  The timescale of
this change -- several years -- is in keeping with the observed
timescales for similar changes in other candidate LBV stars.  Assuming constant
bolometric luminosity with an increase in temperature, observations in
the Rayleigh-Jeans tail of the emission spectrum should follow a
dependence $F_{\nu} \propto \ T^{-3}$.  Thus, the possible drop by
$\sim 0.5$ mag between the observations of \citet{Shri} and our
observations here could indicate a temperature increase in LBV 1806-20
by $\sim 10$\%, with a corresponding decrease in HeI equivalent width.
Also note that if the HeI decrement were due to a temperature {\it
decrease} to $12000-14000K$ (and corresponding radius {\it increase}),
we would expect an apparent K-band {\it brightening} in LBV 1806-20 by
$\sim 1.0$ mag, which is ruled out by the observations.

	In any case, we note that our temperature estimate is based on
the higher equivalent width of HeI (lower temperature), and thus
provides us with a lower estimate on the bolometric luminosity.
Furthermore, the temperature changes hinted at by the HeI variability
and possible associated photometric variability described above are
well within the range of our stated uncertainties in temperature.
Finally, we note that a more precise determination of the temperature
(and thus luminosity) of LBV 1806-20 may be possible in the future,
using detailed models for the stellar spectrum as developed by
\citet{Figer98} for the Pistol Star.

\subsubsection{Distance}

As noted above, the distance to this cluster is the primary subject of
\citet{Corbel} and \citet{CE}, and those papers provide detailed
discussion.  However, two new points are worth emphasizing here.
First of all, the radial velocity of LBV 1806-20 confirms its association
with the molecular clouds at $\sim 4$ kpc or $15.1^{+1.8}_{-1.3}$ kpc.
It is important to note that intermediate distances are essentially
ruled out, as they would require peculiar velocities of LBV 1806-20 as
large as $>100 \ {\rm km \ s^{-1}}$, which is much larger than the
typical velocity dispersions for such massive stars.  Secondly, the
spectral types for stars B, C, and D are all consistent with their
absolute K-band magnitudes if they (and thus LBV 1806-20) are all
located at the distance of $15.1^{+1.8}_{-1.3}$ kpc given by
\citet{CE}.  Furthermore, the observed magnitude of star B combined
with the known range of absolute magnitudes for WC9 stars strongly
rules out the ``near'' distance for this cluster of stars, and
independently sets a minimum distance of at least 9.5 kpc.  Thus,
these stars provide independent confirmation of the original distance
arguments of \citet{Corbel} and \citet{CE}, resulting in a total of
three independent lines of evidence (kinematic distance plus $NH_3$
absorption, absolute magnitude range of Star B, IR extinction) that
show LBV 1806-20 lies at the ``far'' distance of
$15.1^{+1.8}_{-1.3}$..

For the sake of completeness, we also consider the luminosity of Star
A using distance estimates that exclude the kinematic distance
measurement discussed above and in \citet{CE}.  That leaves two major
distance indicators -- the ammonia absorption feature, which places
LBV 1806-20 at a distance of at least $> 5.5$ kpc, and the absolute
K-band magnitude of Star B, which places the cluster at $d>9.5$ kpc.
Note that these two indicators are consistent with each other, as they
are lower limits, and both confirm that LBV 1806-20 does not lie at
the ``near'' kinematic distance of $\sim 4$ kpc.  Taking $d=9.5$ kpc
as a lower limit on the distance to the cluster, we can place lower
limits on the luminosity of LBV 1806-20.  For a temperature at the
minimum of our range above, we have for the lower limit $L_{bol} > 2
\times 10^6 \ L_{\odot}$.  At the upper end of our temperature range
above, the lower limit becomes $L_{bol} > 1.6 \times 10^7 \
L_{\odot}$.  We note again that these estimates ignore the kinematic
information on LBV 1806-20, which places it at $\sim 15$ kpc.  If LBV
1806-20 were in fact to lie at $\sim 10$ kpc, it would have a velocity
deviation of several tens of kilometers per second from the Galactic
rotation curve.  We also note that \citet{lavine} have identified
another WC9 star in this cluster which is $\sim 1$ mag fainter than
Star B, which would increase the ``non-kinematic'' distance lower
limit to $\sim 12.5$ kpc, which is close to the lower end of the
confidence interval for the kinematic distance estimate of \citet{CE}.

We also note here that there is strong reason to believe that all of
these stars (the candidate LBV, Stars B, C, D, and the soft gamma
repeater SGR 1806-20) are in fact members of the same cluster.  The
great rarity of LBV candidates and SGRs in the Galaxy make a chance
association of two so close together very small ($<10^{-5}$
probability \citep{Shri}).  This probability is decreased even
further by the fact that the measured IR extinction towards the LBV
candidate matches the X-ray absorption towards the SGR
\citep{eiken01}.  The additional discovery of a very rare WC9d star
(Star B) and two OB stars (C and D), all with identical reddening,
seems to further show strong evidence for the existence of a physical
association of these massive stars and their remnants.  Additional
support comes from the mid-infrared observations of \citet{Fuchs},
which shows that all of these stars lie within an extended envelope of
mid-IR emission, presumably due to hot dust in their natal molecular
cloud.  The work of \citet{lavine} reveals an additional 2 Wolf-Rayet
stars (one WC9 and one WN5), as well as multiple other OB stars in
this region increases the over-density of massive stars here, which
would seem to cement the conclusion that this is in fact a physical
association of massive stars at the same distance.  In fact, the
surface density of very massive stars here is within a factor of a few
of the most dense concentrations seen in our Galaxy, such as the
Arches cluster \citep{arches}.

Finally, if we exclude both the kinematic arguments above {\it and}
the absolute magnitude of Star B, the ammonia absorption feature
places another lower limit on the distance to the candidate LBV
1806-20 of $>5.7$ kpc \citep{CE}.  This results in a luminosity lower
limit of $>7 \times 10^5 \ L_{\odot}$.  Note again that this distance
is problematic when considering the measured velocity of the molecular
cloud and the candidate LBV, requiring them to have peculiar
velocities of $\sim 100 \ {\rm km s^{-1}}$ compared to the expected
Galactic rotation curve at this location -- an extreme deviation
considering that such massive stars are generally a ``cold''
population in kinematic terms.  Furthermore, at this distance, Star B
would be approximately 4 times less luminous than any WC9 star seen in
the Galactic Center (and the WC9 of \citet{lavine} would be $\sim 12$
times less luminous than any of the Galactic Center WC9 stars).  Thus,
this distance seems incompatible with several observational facts.
Nevertheless, we include this distance estimate here for completeness.

\subsubsection{Is it a cluster or multiple system?}

	Perhaps the strongest caveat to (at least the lower limit for)
the luminosity of LBV 1806-20 is the issue of the star's possible
multiplicity.  A proposed explanation for the observed luminosities of
objects such as the Pistol Star and LBV 1806-20 is that they are in
fact unresolved {\it clusters} of luminous stars.  However, our
speckle observations contradict this conclusion in the case of LBV
1806-20.  For a distance of 15.1 kpc, the upper limit of 60-mas on any
extension corresponds to 0.0044 pc or $\sim 900$ AU FWHM.  This is more
than an order of magnitude smaller than any known cluster.  Thus, it
is very unlikely that LBV 1806-20 is an unresolved cluster of stars.

	On the other hand, OB stars in open clusters are often
($>50\%$ fraction) in binaries, implying that LBV 1806-20 may also be
a binary system.  Our current observations do not strictly rule out a
close binary (or even triple system) as an explanation for LBV
1806-20.  For a circular orbit with a $\sim 450$ AU semi-major axis
and a total binary mass of $\sim 200 M_{\odot}$ (see below), even seen
edge-on, we derive differential orbital velocities for similar-mass
components of $\sim 5-10 \ \rm km \ s^{-1}$ -- well below the observed
line widths.  Thus, it seems quite possible that LBV 1806-20 is a
binary/multiple system.  However, we note that even at our lower
luminosity range, for equal components, each star would be
approximately as luminous as $\eta$ Car.  For mass ratios different
from one, the more massive component would increase in luminosity from
this level.  So, even if LBV 1806-20 is in fact a binary/multiple
system, we are still left with extremely luminous (and massive) stars
composing the system.

Finally, the observed emission line variability and depth of the HeI
$2.112 \mu$m feature also present some possible evidence against
multiplicity in LBV 1806-20.  The emission lines vary by factors of
several, as shown above, implying that one object is probably the
dominant source of radiation in the system.  On the other hand, in an
ultra-luminous binary (such as LBV 1806-20 would have to be),
wind-wind collisions may be a significant source of line emission, and
thus explain the large-amplitude variability in this context.  The
observed depth of the HeI absorption line is less easy to explain
away.  The depth of this line is such that either one component has a
weaker feature and the other star has the deepest such absorption
feature known (to compensate), or both stars have nearly-equally deep
absorption features.  While this latter seems somewhat unlikely, it is
still possible.  Therefore, we conclude that the issue of multiplicity
in LBV 1806-20 remains an open question.

\subsection{The Mass of LBV 1806-20}

	If we assume that LBV 1806-20 is in fact a single star, its
luminosity provides a current mass estimate, assuming that the star
radiates at its Eddington luminosity where radiation pressure at the
surface matches the star's gravity.  At the lower end of the
luminosity range for the best distance estimate ($d = 15.1
^{+1.8}_{-1.3}$ kpc), we derive a minimum present-day mass of $M > 133
\ M_{\odot}$, assuming $N(H)/N(He) = 10$ \citep{STIS}.  Furthermore,
the Eddington limit is a lower limit to the mass -- realistic models
of massive stars typically have $L \sim 0.6-0.7 \ L_{Edd}$
(i.e. \citet{STIS}; \citet{Figer98}).  Thus a realistic lower limit on
the mass of LBV 1806-20 in the same context as similar estimates for
other massive stars is $M > 190 \ M_{\odot}$.  (For completeness, at
the alternative distance lower limits, this mass limit becomes $M > 76
\ M_{\odot}$ ($d > 9.5$ kpc, ignoring kinematic measurements) and $M >
27 \ M_{\odot}$ ($d > 5.5$ kpc, ignoring the absolute magnitudes of
the WC9 stars)).  The following discussion assumes the mass to be $>
190 M_{\odot}$.

	On the other hand, LBV 1806-20 may be a relatively tight
binary, with a projected semi-major axis $<450$ AU (based on our
speckle imaging above).  If so, then the most even distribution of
luminosity (which deviates most from the single star arguments above)
has each star at $>2.5 \times 10^6 \ L_{\odot}$.  Following the same
arguments above based on Eddington luminosities, then each binary
component has a present-day mass $>90 \ M_{\odot}$.  It is not
currently clear whether this scenario is physically plausible.  For
instance, if during formation one of the stars ignited slightly before
the other, the radiation pressure from such a luminous object so
nearby might photoevaporate the second binary component.  (However, as
argued below, shock-induced star formation may invalidate this
argument to some extent).  In any case, our reliance on the Eddington
luminosity for mass estimation indicates that for $L > 5 \times 10^6
\ L_{\odot}$, the total system mass should be $> 190 \ M_{\odot}$.

\subsection{Comparison to the Pistol Star}

The star that most closely resembles LBV 1806-20 in spectral
characteristics (see discussion above) and luminosity is the Pistol
Star \citep{Figer98}.  This is particularly important in that segments
of the astronomical community have been slow to accept the luminosity
estimates for the Pistol Star, and various appeals to exceptional
circumstances in the Pistol Star's unique properties and location
(very close to the Galactic Center) are often made.  However, the
close match in spectral characteristics and luminosity between these
two stars demonstrates that the Pistol Star is not unique in its
properties.  Furthermore, LBV 1806-20 is located at a Galactocentric
radius of $\sim 7$ kpc -- nearly out to the Solar Circle, and
certainly very far removed from the Galactic Center itself.  This
shows conclusively that the formation of very massive stars is an
ongoing process in our Galaxy in the current epoch, and furthermore
that this process is not limited to the extreme environement of the
Galactic Center.

We also note that the similarity between the spectra of LBV 1806-20
and the Pistol Star may indicate that LBV 1806-20 lies at the lower
end of the allowed luminosity range in Figure 8.  However,
verification of this possibility requires higher resolution spectra
and detailed modeling of the stellar atmosphere, which should be
carried out as future work.

\subsection{The Ultimate Fate of LBV 1806-20}

Due in large part to the apparent connection between supernovae and
$\gamma$-ray bursts (i.e. \citet{price}; \citet{hjorth}), there has
been considerable progress in recent years concerning the end-state
evolution of very massive stars.  \citet{heger} present final
evolutionary scenarios for a wide range of stellar masses and
metallicity.  If we assume that LBV 1806-20 had an initial mass of
$\sim 200 \ M_{\odot}$ and has near-solar metallicity, \citet{heger}
show that the star will end its days as a SNIb/c explosion producing a
neutron star.  This result is contrary to the previous ``standard''
assumption that all very massive stars produce black holes (or
disintegrate in pair-instability supernovae), and depends on heavy
mass-loss due to tremendous stellar winds to strip the stellar core
bare and diminish the final core mass.  Note that \citet{gaensler}
infer the presence of just such a strong wind from LBV 1806-20 from
the radio emission of G10.0-0.3.

This result could also extend to the progenitor of SGR 1806-20 -- if
it was only slightly more massive than LBV 1806-20 it could in fact
evolve more rapidly, explode, and yet leave behind a neutron star
remnant.  The fact that SGR 1806-20 is not an ordinary neutron star,
but also a SGR and magnetar \citep{chryssa}, and that other SGRs seem
to be associated with massive star clusters \citep{vrba} may indicate
that the final evolution of some very massive stars not only produces
neutron stars, but in fact tends to produce {\it highly-magnetized}
neutron stars.

\subsection{The Origin of LBV 1806-20 and the Cluster}

	Previously, theorists have argued that stars (and implicitly
tight binaries) greater than $\sim 100 M_{\odot}$ should be impossible
to form under normal circumstances at solar metallicity, due to
radiation pressure on dust grains in the star-forming material,
radiative heating of the accreting gas, and formation of HII regions
before the complete accretion of the envelope \citep{Bond}.  These
calculations assume spherical symmetry, and the likely deviations from
this symmetry (e.g. an accretion disk) will raise the upper limit on
stellar masses.  However, most currently published models of stellar
evolution stop at $\sim 100-120 M_{\odot}$ (i.e. \citet{Leitherer};
\citet{Schaerer}), at least implicitly accepting this as an upper
limit to stellar masses.  Thus, it would seem that the existence of
LBV 1806-20 and of other masssive stars (the Pistol Star, R136a1,
R136a2) may require some revision in our understanding of very massive
star formation, particularly if they are near solar metallicity.
Alternately, pressure-induced star formation due to expanding HII
regions or supernova shocks presents another possible exception to the
upper limit above.

This possibility is particularly intriguing given the environment of
LBV 1806-20 -- the cluster contains both a late-type Wolf-Rayet star
(Star B) as well as at least one neutron star (SGR 1806-20 -- see
\citet{eiken01}).  Since both of these objects are thought to be more
evolved than the candidate LBV (\citet{Massey_review}; \citet{Conti}) and
stellar evolution progresses more rapidly for more massive stars, we
are left to conclude either that their progenitors were even more
massive than LBV 1806-20 (which seems unlikely though not necessarily
impossible -- see below), or that star formation in this location did
not occur at single epoch, but has been spread over time.  SGR 1806-20
alone implies that at least one supernova event must have occurred
previously in this region (though it is not at all certain that this
particular event occurred {\it before} the formation of the candidate LBV star).
As a consequence, it seems quite possible that the formation of LBV
1806-20 was triggered by the expanding HII regions or supernova events
from prior epochs of star formation at this location, resulting in its
unusually high mass.  We also note the simultaneous presence of candidate LBV
and WR stars in the Quintuplet cluster \citep{Figer98}, as well --
confirming that this situation is not unique and revelaing yet another
similarity between LBV 1806-20 and the Pistol Star.

\section{Conclusions}

	We have presented near-infrared imaging and spectroscopy of
the luminous star LBV 1806-20 and 3 other nearby luminous stars.
Based on the results we conclude that LBV 1806-20 has spectral
characteristics very similar to those of AG Car, the Pistol Star, and
P Cyg -- all luminous blue variables -- and is thus likely to be an
LBV itself.  The nearby luminous stars B, C, and D are Wolf-Rayet WC9d
and possible blue hypergiant stars forming part of a cluster which
includes LBV 1806-20.  Their absolute magnitudes and bolometric
luminosities are consistent with other known stars with similar
spectral types, confirming the distance and reddening estimates for
LBV 1806-20 ($15.1 ^{+1.8}_{-1.3}$ kpc and $A_V = 29 \pm 2$ mag).
With a surface temperature in the range 18000-32000K, LBV 1806-20 has
a bolometric luminosity $>5 \times 10^6 \ L_{\odot}$.  If we drop
kinematic measurements of the distance ($15.1 ^{+1.8}_{-1.3}$ kpc), we
have a lower limit on the distance of $>9.5$ kpc, and on the
luminosity of $>2 \times 10^6 \ L_{\odot}$, based on the cluster
stars.  If we drop both the kinematic and cluster star indicators for
distance, an ammonia absorption feature sets yet another lower limit
to the distance of $>5.7$ kpc, with a corresponding luminosity
estimate of $>7 \times 10^5 \ L_{\odot}$ for the candidate LBV
1806-20.  Our speckle imaging shows conclusively that LBV 1806-20 is
{\it not} an unresolved cluster of stars, though it may be a
binary/multiple system.  If LBV 1806-20 is a single or multiple star,
its total mass exceeds $190 M_{\odot}$ (at the $\sim 15$ kpc
distance).  Finally, the presence of LBV 1806-20 with more evolved
stars in the same cluster (i.e. the W-R WCL star and SGR 1806-20)
implies that star formation may have occurred over multiple epochs in
this region of space.

\acknowledgements The authors thank R. Blum and A. Alvarez for help in
obtaining the observations at CTIO, K. Dunscombe, J. Mueller, and
K. Rykoski for assistance at Palomar, and T. Herter for helpful
discussions of these results.  SSE is supported in part at Cornell by
an NSF CAREER award (NSF-9983830), and SGP and DH were partially
supported by this grant.  MAJ was supported at Cornell by a NASA Space
Grant summer research fellowship, and MAG was supported by a NSF REU
fellowship.

\vfill \eject

\clearpage

\begin{deluxetable}{lccc}
\tablecaption{Photometry of Stars}
\startdata 
Star & J-band ($1.25 \mu$m) & H-band ($1.65 \mu$m) & K-band ($2.2 \mu$m) \\ 
\hline 
LBV 1806-20 & 13.93 (8)\tablenotemark{1} & 10.75 (5) & 8.89 (6) \\
B & 17.79 (28) & 13.46 (5) & 10.50 (6) \\
C & 16.38 (8) & 12.76 (5) & 10.80 (6) \\
D & 16.02 (8) & 12.84 (8) & 11.11 (9) \\

\enddata

\tablenotetext{1}{Units are magnitudes. Values in parentheses indicate uncertainties in the final listed digit.}
\end{deluxetable}

\begin{deluxetable}{lccrcl}
\tablecaption{Spectral Lines in LBV1806-20\tablenotemark{1}} 
\startdata Identification & Wavelength ($\rm \AA$) & Centroid ($\rm \AA$) & $W ({\rm \AA})$\tablenotemark{2} & $\Delta V \ (km \ s^{-1})$ \tablenotemark{3} & Comment \\ 
\hline 
HI ($\rm Pa \beta$) & 12817 & 12816 & -82 (5) & 167 & \\
HI (Br 19-4) & 15260 & 15241 & 	-2.0 (9) & \nodata & \\
FeII \& HI (Br 18-4) & 15330, 15341 & 15322 & -4.6 (9) & 197 & Blend \\
HI (Br 17-4) & 15439 & 15425 & -1.1 (6) & \nodata & Broad P-Cygni? \\
HI (Br 16-4) & 15557 & 15550 & -6.9 (5) & \nodata & \\
HI (Br 15-4) & 15701 & 15691 & -6.4 (4) & 189 & \\
FeII\&NIII\&CIII & & 15750 & -7.2 (4) & \nodata & Blend \\
HI (Br 14-4 )& 15880 & 15880 & -7.5 (5) & 168 & \\
FeII & & 15993 &  -1.7 (3) & 115 & \\
HI (Br 13-4) & 16109 & 16110 & -5.4 (3) & 60 & \\
HI (Br 12-4) & 16407 & 16404 & -8.1 (3) & \nodata & Blend \\
FeII	& 1.6435 & 16444 & -7.7 (3) & 178 & \\
FeII & 16768 & 16768 & -6.7 (3) & \nodata & Blend \\
HI (Br 11-4) & 16806 & 16816 & -11.0 (3) & \nodata & Blend \\
FeII & 16873 & 16874 & -8.4 (4) & 210 & Structure? \\
HeI & 17003 & 17009 & -3.5 (3) & 0 & P-Cygni? \\
FeI \& FeII (?)	& & 17110 & -1.60 (16)	& 122? & Blend? \\
HI (Br 10-4) & 17362 & 17365 & -11.7 (2) & 40? & P-Cygni? \\
FeII & 17414 & 17415 & -3.4 (2) & 245? & Blend? \\
FeII & 17449 & 17456 & -1.9 (2) & \nodata & \\
FeII (?) & 20460 & 20460 & -0.7 (3) & & \\
HeI & 20581 & 20581 & -17.4 (3) & 100 & Structure? \\
FeII & 20888 & 20888 & -2.58 (19) & 147 & Blend? \\
HeI & 21121 &  21122 & 0.76 (17) & \nodata & Blend \\
FeII & 21327 & 21321 & -0.97 (12) & \nodata & \\
MgII & 21368 & 21369 & -7.20 (12) & 174 & \\
MgII & 21432 & 21429 & -3.65 (12) & 193 & \\
HI (Br $\gamma$ = Br 7-4) & 21655 & 21655 & -44.2 (3) & 159 & \\
FeII & 21878 & 21870 & -1.14 (11) & \nodata & Blend \\
NaI & 22056 & 22054 & -2.38 (16) & \nodata & \\
NaI & 22083 & 22091 & -0.95 (13) & \nodata & \\
FeII & 22237 & 22240 & -1.22 (10) & \nodata & \\
FeII & 22534 & 22540 & -0.41 (11) & \nodata & \\

\enddata

\tablenotetext{1}{Columns give the line identification, rest
wavelength (in air) , observed line centroid wavelength (in air),
equivalent width, and line full-width at half-maximum converted to
Doppler velocity.}

\tablenotetext{2}{Equivalent widths, with negative values indicating emission and positive values indicating absorption.  Values in parentheses indicate uncertainties in the final listed digit.}

\tablenotetext{3}{Line widths reported result from taking the measured
line width and subtracting the instrumental response function in
quadrature.  The instrumental response was determined from OH sky
lines expected to be free from blending.}

\end{deluxetable}

\clearpage

\begin{figure}
\caption{\it Three-color near-infrared image of the field of LBV
1806-20, coded with J-band = blue, H-band = green, K-band = red.
Labels indicate LBV 1806-20 (A) and the 3 other stars (B,C,D) for
which we obtained near-infrared spectra.  Blue colors indicate
foreground objects, while colors similar to LBV 1806-20 indicate hot
stars with similar reddening.  Coordinates are J2000.0}
\end{figure}

\begin{figure}
\plotone{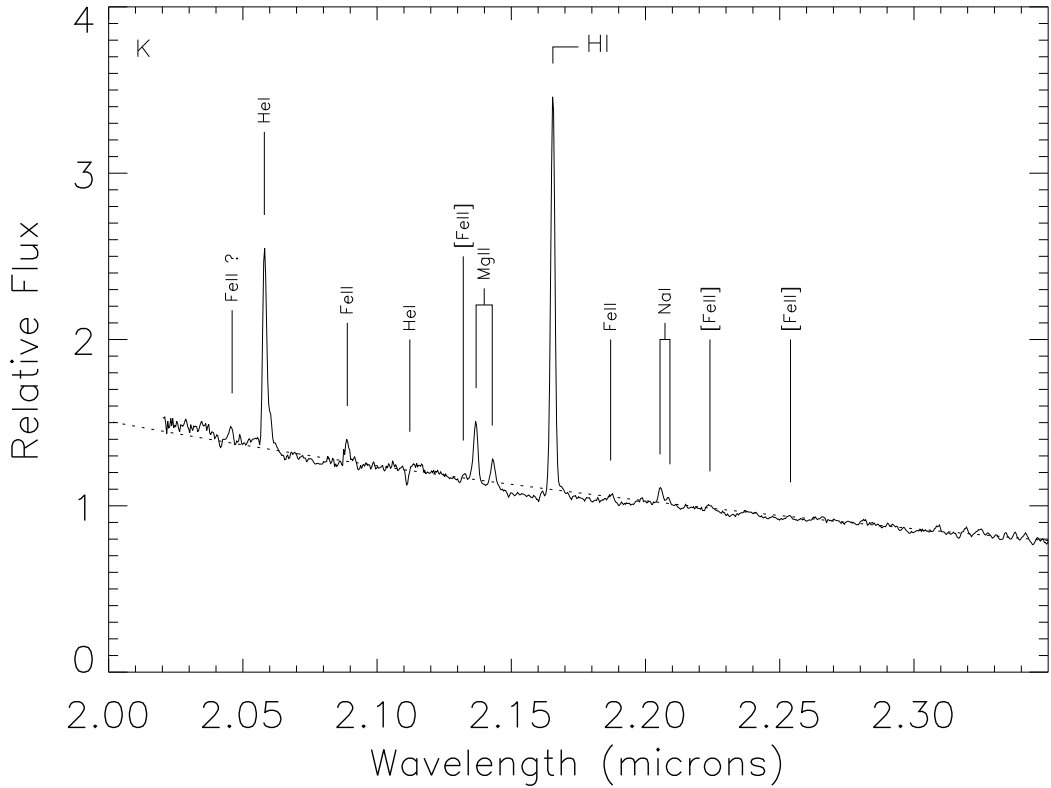}

\caption{\it Near-infrared spectrum of LBV 1806-20 in the K band,
de-reddened with $A_V = 29$ mag, following the reddening law of
\citet{RiekeLebofsky}.  The dotted line indicates the spectral shape
of a blackbody with $T > 12000$ K.  The spectrum closely resembles
that of the Pistol Star and AG Car, and is typical for LBVs at this
wavelength.}
\end{figure}

\begin{figure}
\plotone{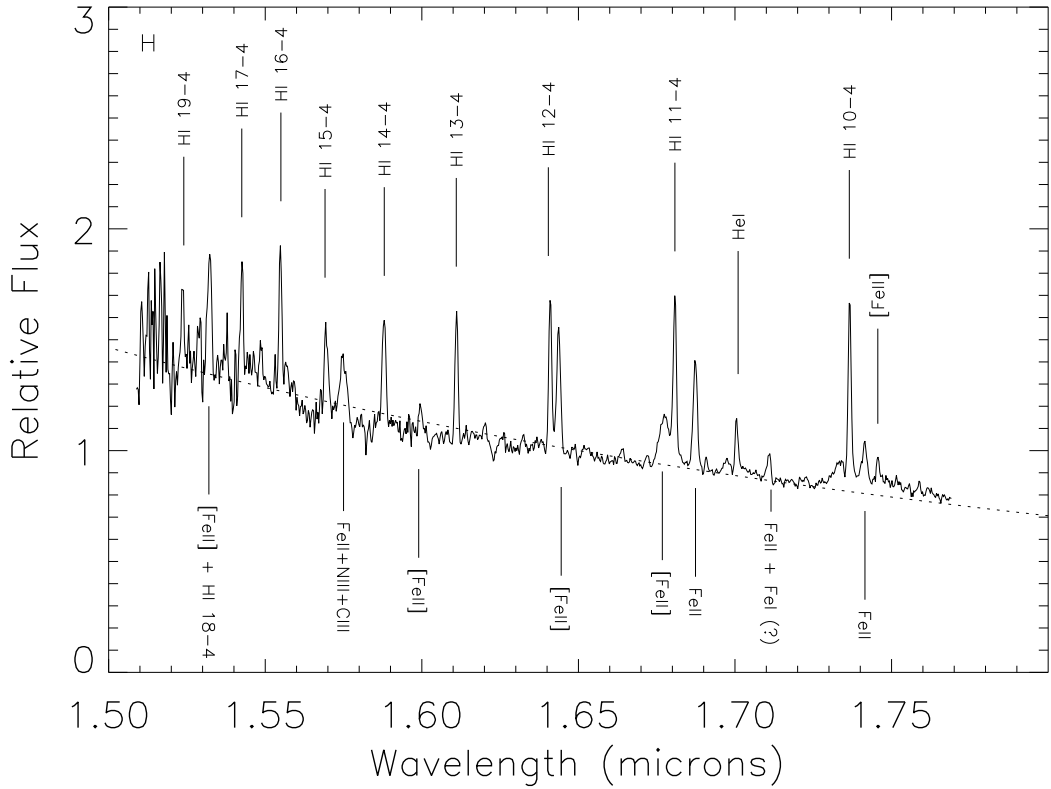}

\caption{\it Near-infrared spectrum of LBV 1806-20 in the H band,
de-reddened with $A_V = 29$ mag, following the reddening law of
\citet{RiekeLebofsky}.  The dotted line indicates the spectral shape
of a blackbody with $T > 12000$ K.  The emission lines are primarily
due to the Brackett series and FeII.}
\end{figure}

\begin{figure}
\plotone{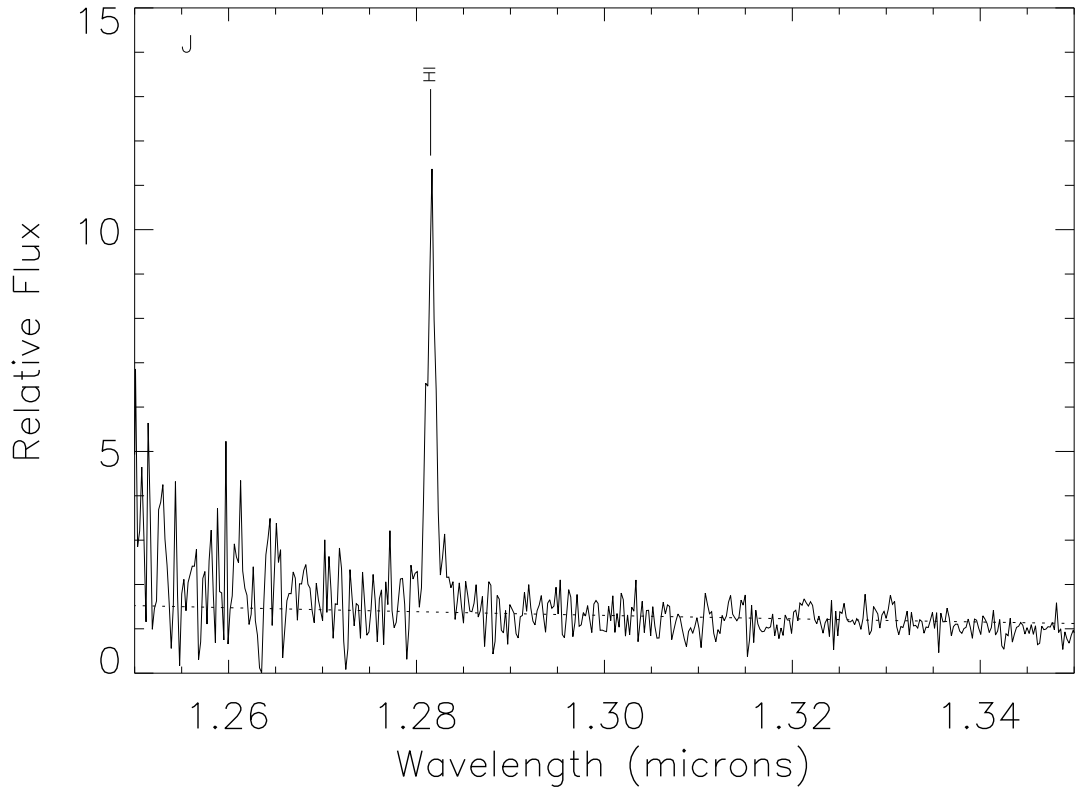}

\caption{\it Near-infrared spectrum of LBV 1806-20 in the J band,
de-reddened with $A_V = 29$ mag, following the reddening law of
\citet{RiekeLebofsky}.  The dotted line indicates the spectral shape
of a blackbody with $T > 12000$ K.  The single strong emission line is
$\rm Pa \beta$.}
\end{figure}

\begin{figure}
\plotone{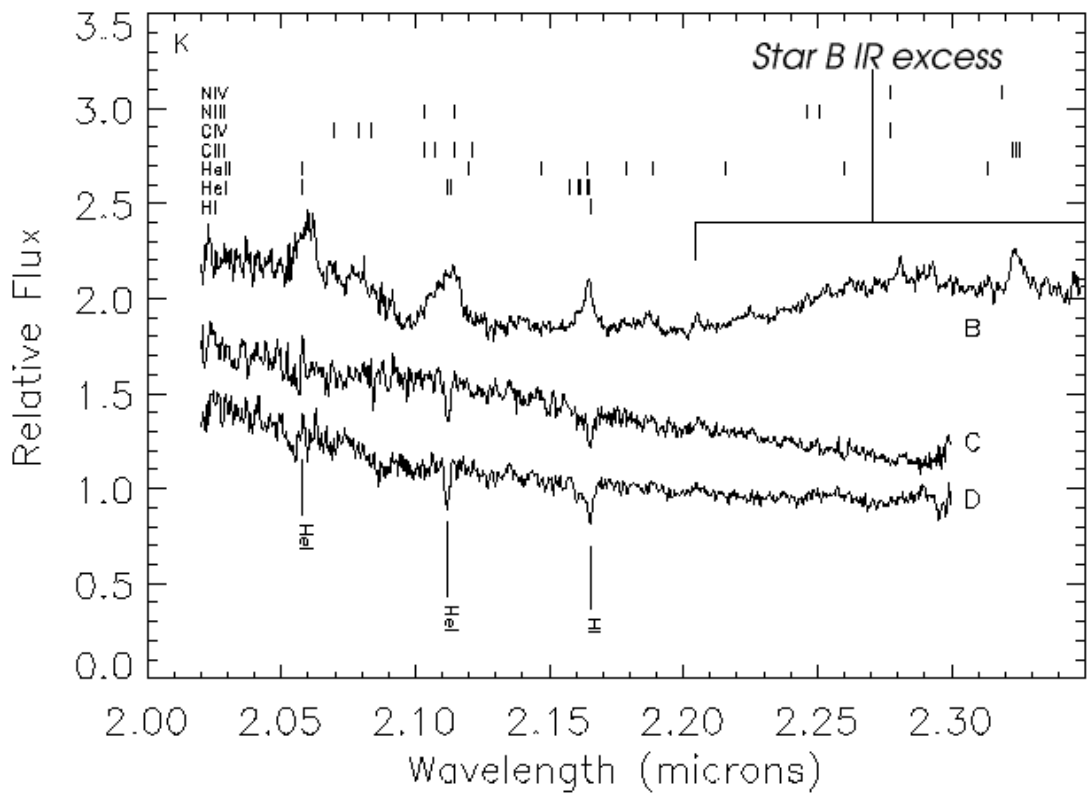}

\caption{\it Near-infrared spectra of stars B, C, and D in the K band.
All spectra have been de-reddened with $A_V = 29$ mag, following the
reddening law of \citet{RiekeLebofsky}, and the spectra of Stars B and
C are vertically shifted for clarity.  Star B shows a bumpy continuum
with a red excess along with broad blended helium, carbon, and
nitrogen emission lines typical of dusty late WC-type Wolf-Rayet
stars.  Stars C and D both show blue continua with HeI $2.112 \mu$m
and HI $\rm ( Br \gamma )$ in absorption, typical of late O- or early
B-type supergiants.  Stars C and D also show some indications of HeI
$2.058 \mu$m in emission/absorption -- possibly a self-absorbed
P-Cygni profile.}
\end{figure}

\begin{figure}
\plotone{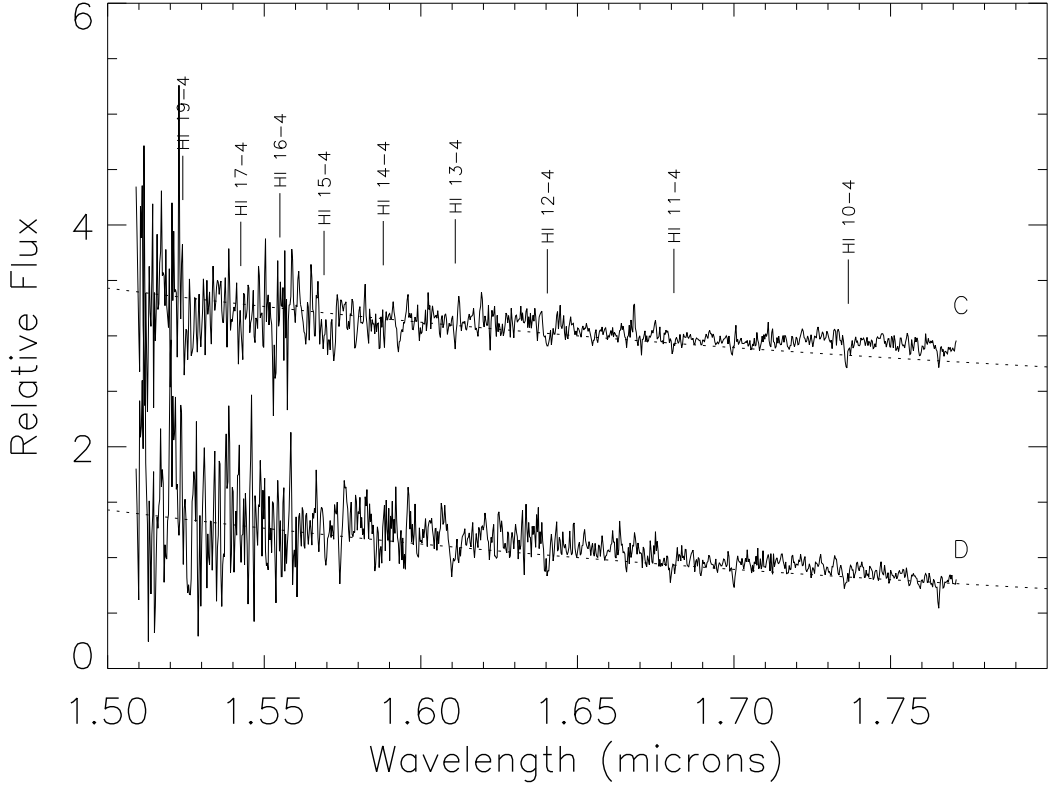}

\caption{\it Near-infrared spectra of stars C and D in the H band,
de-reddened with $A_V = 29$ mag, following the reddening law of
\citet{RiekeLebofsky}.  The dotted lines indicates the spectral shape
of a blackbody with $T > 12000$ K.  The relatively poor spectral
quality is due to a combination of poor observing conditions and the
reddening towards these stars.}
\end{figure}

\begin{figure}
\caption{\it High-resolution speckle images of LBV 1806-20 in the K
($2.2 \mu$m) band taken at the Palomar 5-m telescope on June 30, 1999.
The top left image is of LBV 1806-20 taken with a $0.036 \arcsec$ per
pixel scale ($4.6 \arcsec$ field of view) and a stretch from 0 (sky
background) to the maximum of the image.  This image has a FWHM of
$0.130 \arcsec$, near the telescope diffraction limit of $0.110
\arcsec$.  The top right image is the difference between the image of
LBV 1806-20 and 2 PSF reference stars, with a stretch from 0 (sky
background) to 15 times the RMS noise level ($\sim 0.005$ times the
peak in the top left image).  The bottom image shows the residuals to
a simulated cluster with a Gaussian profile of FWHM = 0.0039 pc, with
the same stretch as the top right image.  Note that residuals for such
a cluster significantly exceed the actual observed residuals,
indicating that LBV 1806-20 is not a cluster of stars.}
\end{figure}

\begin{figure}
{\epsscale{1.0} \plotone{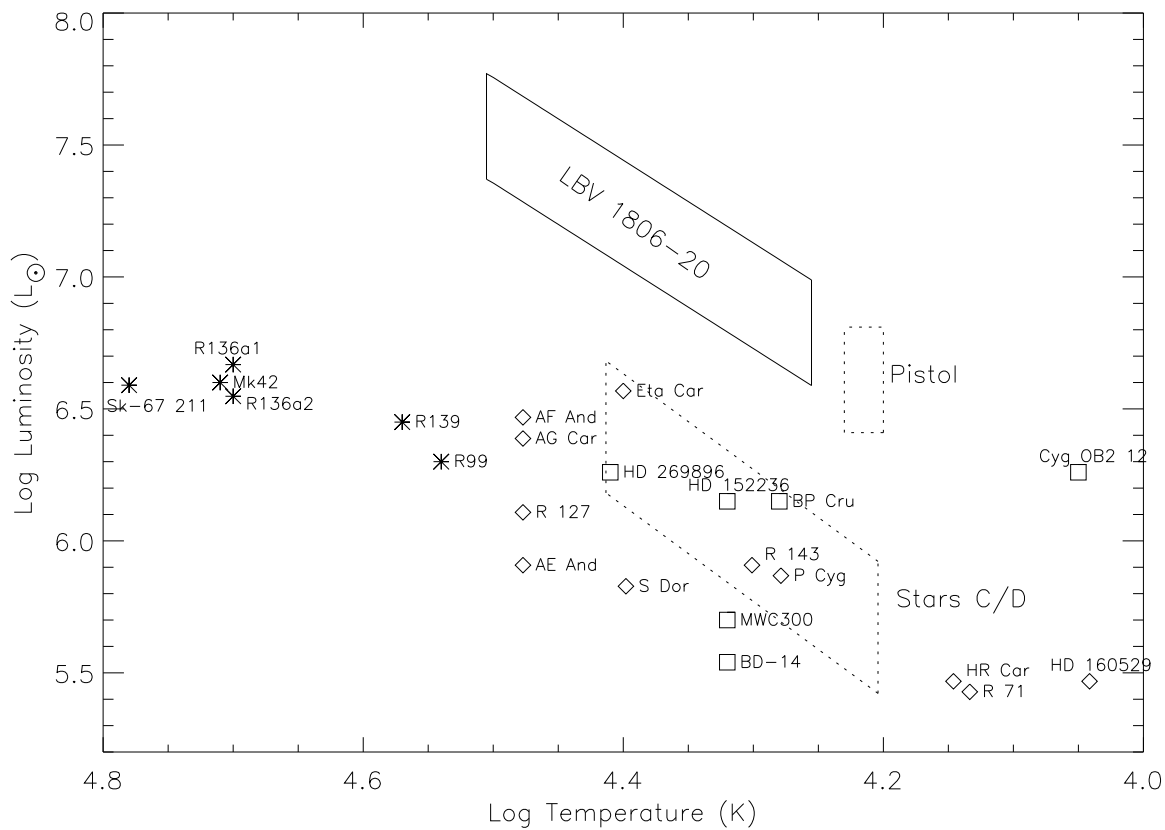}}
\caption{\it Hertzsprung-Russell diagram including LBV 1806-20 and
other hot luminous stars.  Diamonds indicate candidate LBV stars, squares
indicate Ia+ hypergiants, asterisks indicate O-type supergiants, and
regions indicate the locations of LBV 1806-20, Stars C and D, and the
Pistol Star.  Uncertainties in the temperatures of LBV 1806-20 and
Stars C and D dominate the uncertainty in their luminosities.  Note
that even at the lowest temperature, the luminosity of LBV 1806-20
exceeds that of the famous LBV Eta Car and overlaps the upper end of
the Pistol Star's luminosity range.  The allowed
luminosity/temperature range of Stars C and D includes several known
blue hypergiants, further reinforcing their identification as such
stars based on luminosity and spectral features.}

\end{figure}

\end{document}